\documentclass[journal]{IEEEtran}
\usepackage{amsmath}
\usepackage{graphicx}
\usepackage{subcaption}
\usepackage{amsthm}
\usepackage[utf8]{inputenc}
\usepackage[english]{babel}
\newtheorem{theorem}{Theorem}
\usepackage{blindtext}
\usepackage{enumitem}
\usepackage{algorithm}
\usepackage{algpseudocode,algorithm,algorithmicx}

\usepackage{nomencl}
\makenomenclature

\begin{document}
\title{A new state estimation approach-Adaptive Fading Cubature Kalman filter}

\author{Mundla~Narasimhappa,~\IEEEmembership{ Member,~IEEE}
\thanks{His with the Department
of Mechanical Engineering, University of Surrey, Surrey,
GU2 7XH, UK e-mail: (see n.mundla@surrey.ac.uk.}% 
}

% ====================================================================
\maketitle

% === ABSTRACT 
\begin{abstract}
This paper presents a novel adaptive fading cubature Kalman filter (AFCKF) based on double transitive factors. The developed adaptive algorithm is explained in two stages; stage (i) a single transitive factor is used to update the predicted state error covariance, ${\bf \hat P_{k}}^{-}$  based on innovation or residual vector, whereas, in stage (ii), the measurement noise covariance matrix, ${\bf \hat R_{k}^{*}}$ is scaled by another transitive factor. Furthermore, showing the proof concept for estimation of the process noise, ${\bf \hat Q_{k}^{*}}$ and measurement noise covariance matrices by combining the innovation and residual vector in the AFCKF algorithm. It can provide reliable state estimation in the presence of unknown noise statistics. Bench-marking target tracking example is consider to show the performance improvement of the developed algorithms. As compared with existing adaptive approaches, the proposed fading algorithm can provide better estimation results.

\end{abstract}

% === KEYWORDS ====================================================================
% =================================================================================
\IEEEpeerreviewmaketitle

%\mbox{}

\nomenclature{CKF}{Cubature Kalman filter}
\nomenclature{ACKF}{Adaptive Cubature Kalman filter}
\nomenclature{AFCKF}{Adaptive Fading Cubature Kalman filter}
\nomenclature{IAE }{Innovation Based Adaptive Estimation}
%\nomenclature{MMAE }{multiple Model-Based Adaptive Estimation}
\nomenclature{MFF }{Multiple fading factors }
\nomenclature{RMSE }{Root mean squared error}
\nomenclature{E(.)}{Expectation operator} 
\nomenclature{tr(.)}{Trace operator}
% \nomenclature{Q}{Estimated process noise covariance matrix}
% \nomenclature{R}{Estimated measurement noise covariance matrix}

\printnomenclature

% === I. INTRODUCTION =============================================================
% =================================================================================
\section{Introduction}

\IEEEPARstart{D}{iff} variants of nonlinear state estimators have been developed in the literature; Extended Kalman Filter (EKF) \cite{kim2009stability}, Unscented Kalman filter (UKF) \cite{soken2009adaptive} and cubature Kalman filter (CKF) \cite{arasaratnam2009cubature} are the most popular methods. In the EKF, the non-linear function is approximated through Taylor or Jacobian calculations \cite{guo2020robust}, whereas in the UKF \cite{wan2000unscented}  has been developed based on the sigma points approach. However, UKF has afford considerably accurate estimation  than the EKF estimator. But, the estimation accuracy of the UKF is limited for higher-order systems analysis. The CKF \cite{arasaratnam2009cubature} can be developed  and  being widely applied into various real world estimation problems in \cite{bar2004estimation, yue2019novel,singh2020exponentially, li2020application, kardan2018improved, pramanik2019accurate, wang2020generalized}. 

System and measurement noise models may not known exactly and even varying with time in practice. However, the filter  becomes divergence and the performance can be degraded. In \cite{mohamed1999adaptive}, standard adaptive approaches have been developed in the literature; such as (i) Innovation Based Adaptive Estimation (IAE) and (ii) Residual Based Adaptive Estimation (RAE).(iii) multiple Model-Based Adaptive Estimation (MMAE) \cite{liu2020adaptive}. In adaptive estimation \cite{zhao2015design},  estimation of the noise covariance matrices were developed neither residual nor innovation sequence in the first and second method, whereas, in the third approach, several filters are running in parallel, however,  it causes an increment in the storage burden \cite{zhou2019new}. To realize in the MMAE, type of distribution of the innovation or residual vector must be known within a window for all epochs\cite{kim2009adaptive}. System uncertainty impact on filter performance and sub-optimal when the  noise covariance matrices are varied under the fault measurements\cite{yue2019novel,liu2020adaptive}. By introducing the  time-dependent variable, called  fading factor, named it as Adaptive Fading Kalman Filter (AFKF) \cite{ xia1994adaptive}. Nonlinear fading EKF \cite{kim2009stability, weixi2011multiple}, and adaptive fading  AFUKF\cite{li2017stochastic}, adaptive fading cubature KF \cite{guo2020robust, zhao2015design}. The ACKF with a single fading factor was developed for estimating the noise covariance matrices. Technically, by using multiple fading factors (MFF) in the ACKF filter has more beneficial than the single fading factor-based AFKF.  AFUKF \cite{guo2020robust} and AFCKF have the same performance, except  in higher dimensional system estimation. The proposed methodology is developed based on double transitive factors and is the aim of this paper, that can correct the gain may be utilized by varying the predicted and measurement noise covariances. Best of author knowledge, there is a limited contribution on double transitive factors based nonlinear adaptive fading CKF algorithm. The attributions of this paper have the follows;  This paper investigates the nonlinear state estimation problem by applying the adaptive fading cubature Kalman filter. An adaptive fading CKF framework is developed with double transitive factors. The state error covariance is adapted with a single transitive factor, then the process noise covarinace matrix adaption is made by adjusting the properties of nonlinear systems.  Another transitive factor is developed in this paper for scaling with the measurement noise covariance matrix adaption. Then, these errors are assumed to be non-Gaussian rather than satisfying the Gaussian distribution.A bench marking tracking example is shown to assess the performance improvement of the developed algorithm.
The rest of the paper is organized as follows; the adaptive fading cubature Kalman filter based on double transitive factors is presented in Section II. In section III is shown the tracking example for validating proposed algorithms. The conclusion of the paper is presented in  Section IV.

\section{Adaptive fading cubature Kalman filter based on double transitive factors}
\subsection{Problem formulation}
Considering the nonlinear discrete-time stochastic system, and  measurement equations are

\begin{equation}\label{eqn.1}
 {\bf x_{k}}={\bf f({x_{k-1}})}+{\bf w_{k-1}}
 \end{equation}
\begin{equation}\label{eqn.2}
 {\bf z}_{k}={\bf h({\bf x_{k}})}+{\bf v_{k}}
 \end{equation}

where, state vector, $\bf x_{k}  \in R^{n}$ with n- dimension and ${\bf f({x}_{k-1})}$ is the nonlinear function for the state, ${\bf u_{k}}  \bf \in R^{r}$ is considerable control input,  and  ${\bf w_{k-1}} \sim \mathbf{N}(0, {\bf Q_{k}})$ the the process  is assumed to be Gaussian. the  measurement  vector ${\bf z_{k}} \bf \in R^{m}$ at time ${k}$. The measurement nonlinear function $  {\bf h({x}_{k})}$ and the measurement noise ${\bf v_k} \sim \mathbf{N}(0, {\bf R_{k}})$ is also assumed to be Gaussian. Aim of this paper is develop a novel adaptive fading CKF based on double transitive factors. The recursive solution to the AFCKF algorithms for estimating noise statistics. The transitive factor, ($a_{1}$) for updating the predicted state error and  double transitive factor, ($a_{2}$) for measurement noise covarinace matrix. Furthermore, these covariance matrices are estimated by using innovation and residual vector difference. The detail derivations are explained in the following sub-sections \ref{section:A} and \ref{section:B}.
\subsection{Adaptive fading CKF Scheme(AFCKF) for P-adaption}
\label{section:A}
% In this section, an innovative sequence is used to develop the adaptive fading CKF scheme. During the filter process, the innovation sequence is the additional information to the filter. Moreover, the innovation and residual sequences are combined and are used to estimate the process noise covariance matrix, $Q-{k}$.

In this section, innovative or residual sequence \cite{mohamed1999adaptive} is used to develop the adaptive fading CKF scheme for the predicted state error covariance  adaption with defined single transitive factor, simply named it as AFCKF-P adaption algorithm. The innovation sequence is the difference between the measurement, ${\bf z_{k}}$ and predicted measurements, $\bf {h} ({\bf \hat z_{k}}^{-})$ , is defined as
\begin{equation}
{\bf \upsilon_{k}}={\bf z_{k}}-\bf {h} ({\bf \hat z_{k}}^{-})
\label{eq:3}
\end{equation}
By considering the measurement equation into equation (\ref{eq:3}) and  applying the  expectation on both sides,  we can get the auto-covariance of the innovation sequence is
 \begin{equation}
  \label{eq:23}
  \begin{aligned}
E[{\bf \upsilon_{k}} {\bf \upsilon_{k}}^\top]&=E[{\bf h}({\bf {x}_{k}}-{\bf \hat {x}^{-}_{k}})+{\bf v_k}] [{\bf  h}({\bf {x}_{k}}-{\bf \hat {x}^{-}_{k}})+{\bf v_k})]^\top\\
&= E[{\bf h}({\bf \Delta \hat {x}^{-}_{k}})+{\bf v_k}] [{\bf  h}({\bf \Delta \hat {x}^{-}_{k}})+{\bf v_k})]^\top\\
&= {\bf  h}E[({\bf \Delta \hat {x}^{-}_{k}}) ({\bf \Delta \hat {x}^{-}_{k}})^\top] {\bf  h^\top} + E[ {\bf v_k} {\bf v}^\top_k] \\
&= {\bf h}{\bf \hat P_{k}} {\bf  h^\top} + {\bf  R_{k}} \\
&=\sum\limits_{i=0}^{2L}{W_{i}^{c}}({({\bf Z_k})_i}-{\bf \hat z_{k}}^{-})[{({\bf Z_k})_i}-{\bf \hat z_{k}}^{-})^{T}+  \bf R_{k}\\
&={ \bf C_{\bar \upsilon k}}
\end{aligned}
\end{equation} 

% \begin{equation}
% %\begin{split}
% {\bf  P_{zz, k}}={ \bf C_{\bar \upsilon k}}=\sum\limits_{i=0}^{2L}{W_{i}^{c}}({({\bf Z_k})_i}-{\bf \hat z_{k}}^{-})[{({\bf Z_k})_i}-{\bf \hat z_{k}}^{-})^{T}+ \\ \bf R_{k}
% \label{stateSpaceForm1}
% %\end{split}
% \end{equation}
where, ${ \bf C_{\bar \upsilon k}}$ is the theoretical covariance matrix of the innovation sequence, ${\bf {x}_{k}}$  and ${\bf {x}_{k}}$ are the predicted and estimated states. ${W_{i}^{c}}$ is the weights, L is the dimension of the state vector, ${\bf \Delta \hat {x}^{-}_{k}}$ error between actual and estimated states. T is the transpose. 
Furthermore, the estimated covariance matrix of innovation through windowing average method \cite{mohamed1999adaptive} is
\begin{equation}
{\bf \hat C_{\bf \bar \upsilon k}}=\frac {1}{N_{w}}\sum_{j=j0}^{k}{\bf \bar \upsilon}_{j}{\bf \bar \upsilon}_{j}^T
% \hat{x}	\hat{x}
\label{stateSpaceForm1}
\end{equation}
where $N_w$ is the moving window width. As it is known, when the sample window, $N_w$ increases the sample covariance tends to close to the actual value. However, it satisfies the stationary processes, whereas in the non-stationary conditions, the true innovation-covariance matrix varying with time at each epoch.  The  selection of transitive factor $ {\bf a_{1}}(k)$ is evaluated as
\[
 {\bf a_{1}}(k)=\left\{ \begin{array}{cc}
  1, &  if\  {tr({\bf \hat C_{\bf \bar\upsilon k}})}>{tr({\bf \hat P}_{\bf \bar \upsilon k})}.\\
  \frac{tr({\bf \hat C_{\bf\bar \upsilon k}}-\bf R_{k})}{tr(({\bf  P_{zz,k}}-\bf R_{k}))}, & { otherwise}.
\end{array} \right.
\]
with  ${\bf a_{1}}(k)$ being a adaptive transitive factor, this value is varied accord to the actual and estimated covariance matrices of the innovation vector, $ {\bf  \upsilon_{k}}$. Here,  $tr$ is the trace function. If the   are larger than that of actual values, ${\bf a_{1}}(k)$. Otherwise, ${\bf a_{1}}(k)$ is approximately calculated the ratio difference of estimated covariance matrix of the innovation sequence and measurement noise covariance matrix. 

% and ${{\bf \hat P}_{\bf \bar \upsilon k}}$  is the estimated covariance matrix of residual sequence  expressed as
% \begin{equation}
% {{\bf \hat P}_{\bf  \upsilon k}}={\bf  \upsilon_{k}\bf \upsilon_{k}^T}
% \label{eqn:3}
% \end{equation}

The predicted state covariance ${\bf \hat P_{k}}^{-}$  is scaled by a single transitive factor is \cite{narasimhappa2016arma}
\begin{equation}
{\bf \hat P_{k}}^{-}= {\frac{1}{\bf a_{1}(k)}}{\bf \hat P_{k-1}}^{-}
\label{eqn:4}
\end{equation}

Generally the predicted  state error covariance  as in (\ref{eqn:4}) and is evaluated from the innovation samples. The ${\bf \hat P_{k}}^{-}$ should satisfy the symmetric and positive semi-definite at each time step. The state error covariance is adaptively updated with the transitive factor and then it can be used into the  Q adaption (see in Theorem 1). In this stage,  error variation in the state models due to the fault measurements, thus, the AFCKF-prediction equation is inaccurate, and causing to increase the error covariance and consequently of Kalman gain. Therefore, introducing another adaptive transitive factor in the following subsection, which takes to quickly react to the state variation.

% As it is known, when the sample window, $N_w$ increases the sample covariance tends to close to the actual value. However, it satisfies the stationary processes, whereas in the non-stationary conditions, the true innovation-covariance matrix varying with time at each epoch. In this situation, we known the exact knowledge of the observation matrix and measurement noise covariance, respectively.

\subsection{Adaptive fading CKF Scheme(AFCKF)-R adaption }
\label{section:B}
Once we update the optimal transitive factor in stage I, then it can forward to second stage II. The AFCKF algorithm,  the measurement noise covarinace matrix is scaled by the double transitive factor, named it as AFCKF-R adaption. The transitive factor ${\bf a_{2}({k})}$ is evaluated as

\[
 {\bf a_{2}(k)}=\left\{ \begin{array}{cc}
  1, & {\bf \hat P_{\bf \upsilon k}}>{tr({\bf \hat C_{\bf \bar \upsilon k}})}.\\
  \frac{tr({\bf \hat C_{\bf \bar \upsilon k}})}{tr({ \bf C_{\bf \bar \upsilon k}})}, & { otherwise}.
\end{array} \right.
\]

where ${\bf \hat P_{\bf \eta k}}= {{\bf  \eta}_{k} {\bf \eta }_{k}^T} $ is the estimated covariance matrix of the residual sequence. The double transitive factor ${\bf a_{2}(k)}$ is used to multiply with the measurement noise covariance matrix. Thus, equation (\ref{eqn:7}) can be written as
\begin{equation}
%\begin{split}
{\bf  P_{zz, k}}=\sum\limits_{i=0}^{2L}{W_{i}^{c}}[{({\bf Z_k})_i}-{\bf \hat z_{k}}^{-}][{({\bf Z_k})_i}-{\bf \hat z_{k}}^{-}]^{T}+{\bf a_{2}}(k){\bf R_{k}}
\label{eqn:7}
%\end{split}
\end{equation}

The relation between $ {\bf a_{2}({k})}$ and $\bf R_k$ are proportional. If $ {\bf a_{2}({k})}$ is large,  $\bf R_k$ also becomes larger, which means the Kalman gain is less and then the influence of uncertainty is more we can trust more on the measurements, other vice verse. \cite{narasimhappa2016arma}. In the section A  (AUFKF-P adaptation algorithm) predicted state error covariance and B (AUFKF-R adaptation algorithm) measurement noise covariance matrix are adapted with  double factor and is given in theorem 1.

% Theorem 2 presents the detailed derivation for estimating the measurement noise covariance matrix is given in the following section.

% In this paper, two theorems are developed and shown proper proof  for the process and measurement  noise covariance matrix estimators. as presented in the following subsection. 

\begin{theorem}
Suppose the noise statistics of the process and measurement noise parameters are very small within a considerable  window size  of  ${N_w}$. Then, a novel process noise statistic estimators are estimated with difference of innovation and residual vector and transitive factor, and then  $a_{1}(k)$ in the predicted state error covariance,   $Q_{k}$ is developed. Moreover, $a_{2}(k)$ in the measurement noise covariance is estimated, $R_{k}$ are derived as
 \begin{equation}
  \label{eqn:8}
  \begin{aligned}
 {\bf h} {\bf \hat Q}_{k-j}^{*} {\bf  h^\top} &= \frac{1}{N_{w}} \sum\limits_{j=1}^{N_{w}}  ({\bf \upsilon_{k-j}}-{\bf \eta_{k-j}}) ({\bf \upsilon_{k-j}}-{\bf \eta_{k-j}})^\top \\ &\quad - {\bf h} (\frac{1}{2L}\sum\limits_{i=1}^{2L}{({\bf X_{k-1}})_i} {({\bf X_{k-1}})_i^\top} -{\bf \hat x_{k-1}}{\bf \hat x_{k-1}}^{{-}^\top}){\bf  h^\top} \\ &\quad + {\bf h}{\bf \hat P_{k}} {\bf  h^\top} \\
 {\bf \hat R}_{k-j}^{*}  &=\frac{1}{{N_w}}\sum\limits_{j=0}^{N} {\bf \upsilon_{j}}  {\bf \upsilon_{j}}^\top-{\bf h_{k-j}} {\bf \hat P_{k-j}}  {\bf h_{k-j}^\top} 
    % {\bf \hat P_{k-1}^{-}} &=\frac{1}{a_{1}(k)}\{\frac{1}{2L}\sum\limits_{i=1}^{2L}{({\bf X_{k-1}})_i} {({\bf X_{k-1}})_i^\top} -{\bf \hat x_{k-1}}{\bf \hat x_{k-1}}^{{-}^\top}\} \\&\quad + {\bf \hat Q}_{k-j}^{*}\\
\end{aligned}
\end{equation}
\end{theorem}

\begin{proof} Let us assume that  the process and measurement covariances of noise statistics are varies from $t_{k-N_{w}}$ to $  t_{k}$, the window width, ${N_w}$ and there are ${N_w}$ measurements. The innovation and residual sequence are represented already in equation (9) and (10).  By defining  predicted and estimated state errors \cite{narasimhappa2021} are 
 \begin{equation}
  \label{eqn:11}
  \begin{aligned}
{\bf \Delta \hat {x}^{-}_{k}}&={\bf {x}_{k}}-{\bf \hat {z}^{-}_{k}},\\
{\bf \Delta \hat {x}_{k}}&={\bf {x}_{k}}-{\bf \hat {x}_{k}}
  \end{aligned}
\end{equation} 
%  \begin{equation}
%  {\bf \upsilon_{k}}={\bf z_{k}}-{\bf  h}({\bf \hat x_{k}^{-}})
%   \label{eqn:9}
%   \end{equation}
%  and
%   \begin{equation}
%  {\bf \eta_{k}}={\bf z_{k}}-{\bf  h}({\bf \hat x_{k}})
%   \label{eqn:8}
%   \end{equation}
% By considering the measurement equation in (\ref{eqn.2}) and substitute  into  in the innovation and residual sequence (\ref{eqn:7}), then we have to get \cite{narasimhappa2021}
%  \begin{equation}
%   \label{eqn:9}
%   \begin{aligned}
%  {\bf \upsilon_{k}}&={\bf z_{k}}-{\bf  h}({\bf \hat z_{k}^{-}})\\
% &= {\bf {\bf  h}({\bf x_{k}})}+{\bf v_{k}}-{\bf  h}{\bf \hat x_{k-1}^{-}}\\
% &= {\bf h}({\bf {x}_{k}}-{\bf \hat {x}^{-}_{k}})+{\bf v_k}
% \end{aligned}
% \end{equation}  
  
%  and 
 
%   \begin{equation}
%   \label{eqn:10}
%   \begin{aligned}
%  {\bf \eta_{k}}&={\bf z_{k}}-{\bf  h}({\bf \hat x_{k}})\\
% &= {\bf {\bf  h}({\bf x_{k}})}+{\bf v_{k}}-{\bf  h}{\bf \hat x_{k}}\\
% &={\bf h}({\bf {x}_{k}}-{\bf \hat {x}^{-}_{k}}) +{\bf v_{k}}\\
% \end{aligned}
% \end{equation}

% where, innovation sequence,  ${\bf h}({\bf {x}_{k}}-{\bf \hat {x}^{-}_{k}}) $ is equal to  $({\bf z_{k}}-{\bf \hat z_{k-1}^{-}})$.
% By considering the  equation (\ref{eqn.2}), the residual sequence is obtained thorough measurement and estimated. Once we substitute the measurement in the equations, state and nonlinear functions. Thus, it satisfies non-orthogonal condition. 

As per expectation and correlation definition, if  the process and measurement noises are uncorrelated, then $E [({\bf \Delta \hat {x}^{-}_{k}}) {\bf v_{k}}^\top] =0 $. We can also check cross-correlation of  $E [({\bf \Delta \hat {x}^{-}_{k}}) ({\bf \Delta \hat {x}_{k}})^\top] $ and  $[{\bf v_{k}} ({\bf \Delta \hat {x}^{-}_{k}})^\top]$, we have 
%
%  \begin{equation}
%   \label{eq:23}
%   \begin{aligned}
% E[({\bf \Delta \hat {x}^{-}_{k}}) ({\bf \Delta \hat {x}_{k}})^\top]
% &=E\{ [({\bf {x}_{k}}-{\bf \hat {x}^{-}_{k}})] [({\bf {x}_{k}}-{\bf \hat {x}_{k}})]^\top\}\\
% &=E[({\bf {x}_{k}}-({\bf \hat {x}_{k}}-{\bf K_{k}} {\bf \upsilon_{k}})] \\& [({\bf {x}_{k}}-({\bf \hat {x}^{-}_{k}}+ {\bf K_{k}} {\bf \upsilon_{k}})]^\top \\
% &= E[({\bf {x}_{k}}-{\bf \hat {x}^{-}_{k}})-{\bf K_{k}} {\bf \upsilon_{k}})] \\& [({\bf {x}_{k}}-{\bf \hat {x}_{k}})+ {\bf K_{k}} {\bf \upsilon_{k}})]^\top \\
% \end{aligned}
% \end{equation}

%
 \begin{equation}
  \label{eq:24}
  \begin{aligned}
E[{\bf \eta_{k}} {\bf \eta_{k}}^\top]&=E[{\bf h}({\bf {x}_{k}}-{\bf \hat {x}_{k}}) +{\bf v_{k}}] [{\bf h}({\bf {x}_{k}}-{\bf \hat {x}_{k}}) +{\bf v_{k}}]^\top\\
&= E[{\bf h}({\bf \Delta \hat {x}_{k}})+{\bf v_k}] [{\bf  h}({\bf \Delta \hat {x}_{k}})+{\bf v_k})]^\top\\
&= {\bf  h}E[({\bf \Delta \hat {x}_{k}}) ({\bf \Delta \hat {x}_{k}})^\top] {\bf  h^\top} - E[ {\bf v_k} {\bf v}^\top_k] \\
&= {\bf  R_{k}}-{\bf h}{\bf \hat P_{k}} {\bf  h^\top}  \end{aligned}
\end{equation}

By considering the cross correlation between residual and innovation sequence. Then, we can rewrite equation are
 \begin{equation}
  \label{eq:23}
  \begin{aligned}
E[{\bf \eta_{k}} {\bf \epsilon_{k}}^\top]&=E[{\bf h}({\bf {x}_{k}}-{\bf \hat {x}^{-}_{k}})+{\bf v_k}] [{\bf  h}({\bf {x}_{k}}-{\bf \hat {x}_{k}})+{\bf v_k})]^\top\\
&= E[{\bf h}({\bf \Delta \hat {x}^{-}_{k}})+{\bf v_k}] [{\bf  h}({\bf \Delta \hat {x}_{k}})+{\bf v_k})]^\top\\
&= {\bf  h}E[({\bf \Delta \hat {x}^{-}_{k}}) ({\bf \Delta \hat {x}_{k}})^\top] {\bf  h^\top}+ {\bf  h} E[({\bf \Delta \hat {x}^{-}_{k}}) ({\bf v_k})^\top] \\&+ {\bf  h}^\top E[({\bf \Delta \hat {x}^{-}_{k}}^\top) ({\bf v_k})]+ E[ {\bf v_k} {\bf v}^\top_k] \\
\end{aligned}
\end{equation} 
 
%  \begin{equation}
%   \label{eq:23}
%   \begin{aligned}
% E[{\bf \upsilon_{k}} {\bf \eta_{k}}^\top]&=E[{\bf h}({\bf {x}_{k}}-{\bf \hat {x}^{-}_{k}})+{\bf v_k}] [{\bf  h}({\bf {x}_{k}}-{\bf \hat {x}_{k}})+{\bf v_k})]^\top\\
% &= E[{\bf h}({\bf \Delta \hat {x}^{-}_{k}})+{\bf v_k}] [{\bf  h}({\bf \Delta \hat {x}_{k}})+{\bf v_k})]^\top\\
% &= {\bf  h}E[({\bf \Delta \hat {x}^{-}_{k}}) ({\bf \Delta \hat {x}_{k}})^\top] {\bf  h^\top}+ {\bf  h} E[({\bf \Delta \hat {x}^{-}_{k}}) ({\bf v_k})^\top] \\&+ {\bf  h}^\top E[({\bf \Delta \hat {x}^{-}_{k}}^\top) ({\bf v_k})]+ E[ {\bf v_k} {\bf v}^\top_k] \\
% \end{aligned}
% \end{equation} 

By taking the difference between innovation and residual and then applying the expectation for them is as follows that Equation (18) and (21), the innovation and residual covariance, and substitute the predicted state error covarinace inside the equations, rewrite the equation  as

%  \begin{equation}
%   \label{eq:t}
%   \begin{aligned}
% E[({\bf \upsilon_{k}}-{\bf \eta_{k}}) ({\bf \upsilon_{k}}-{\bf \eta_{k}})^\top]&=E[{\bf \upsilon_{k}} {\bf \upsilon_{k}} ^\top] + E[{\bf \eta_{k}} {\bf \eta_{k}} ^\top]  \\ & \quad +E[{\bf \upsilon_{k}} {\bf \eta_{k}} ^\top]+E[{\bf \eta_{k}} {\bf \upsilon_{k}} ^\top]\\
% \end{aligned}
% \end{equation}

 \begin{equation}
  \label{eq:t}
  \begin{aligned}
E[({\bf \upsilon_{k}}-{\bf \eta_{k}}) ({\bf \upsilon_{k}}-{\bf \eta_{k}})^\top]&=E[{\bf \upsilon_{k}} {\bf \upsilon_{k}} ^\top] + E[{\bf \eta_{k}} {\bf \eta_{k}} ^\top] \\ &\quad +E[{\bf \upsilon_{k}} {\bf \eta_{k}} ^\top]+E[{\bf \eta_{k}} {\bf \upsilon_{k}} ^\top]\\
&= {\bf h}{\bf \hat P_{k-1}} {\bf  h^\top} + {\bf  R_{k}} + {\bf  R_{k}}- {\bf h}{\bf \hat P_{k}} {\bf  h^\top} \\ &\quad -2({\bf h}{\bf \hat P_{k}} {\bf  h^\top} + {\bf  R_{k}} {\bf h}{\bf {K_{k}^\top}} {\bf  h^\top}+ {\bf  R_{k}} )  \\
&= {\bf h}{\bf \hat P_{k-1}} {\bf  h^\top}- {\bf h}{\bf \hat P_{k}} {\bf  h^\top} \\
&= {\bf h} (\frac{1}{2L}\sum\limits_{i=1}^{2L}{({\bf X_{k-1}})_i} {({\bf X_{k-1}})_i^\top} \\ &\quad -{\bf \hat x_{k-1}}{\bf \hat x_{k-1}}^{{-}^\top} +{\bf Q_{k-1}}) {\bf  h^\top} - {\bf h}{\bf \hat P_{k}} {\bf  h^\top} \\
\end{aligned}
\end{equation}
We can separate out the process noise covarinace matrix from the above equations.  
 \begin{equation}
  \label{eq:14}
  \begin{aligned}
 {\bf h} {\bf Q_{k-1}} {\bf  h^\top} &= E[({\bf \upsilon_{k}}-{\bf \eta_{k}}) ({\bf \upsilon_{k}}-{\bf \eta_{k}})^\top] \\ &\quad - {\bf h} (\frac{1}{2L}\sum\limits_{i=1}^{2L}{({\bf X_{k-1}})_i} {({\bf X_{k-1}})_i^\top} -{\bf \hat x_{k-1}}{\bf \hat x_{k-1}}^{{-}^\top}){\bf  h^\top} \\ &\quad + {\bf h}{\bf \hat P_{k}} {\bf  h^\top} 
\end{aligned}
\end{equation}

On the other hand, the above equation can be  approximate the limited number of sample for the  difference of innovation and residual sequence is 
 \begin{equation}
  \label{eq:t}
  \begin{aligned}
 {\bf h} {\bf Q_{k-1}^{*}} {\bf  h^\top} &= \frac{1}{N_{w}} \sum\limits_{j=1}^{N_{w}}  ({\bf \upsilon_{k-j}}-{\bf \eta_{k-j}}) ({\bf \upsilon_{k-j}}-{\bf \eta_{k-j}})^\top \\ &\quad - {\bf h} (\frac{1}{2L}\sum\limits_{i=1}^{2L}{({\bf X_{k-1}})_i} {({\bf X_{k-1}})_i^\top} -{\bf \hat x_{k-1}}{\bf \hat x_{k-1}}^{{-}^\top}){\bf  h^\top} \\ &\quad + {\bf h}{\bf \hat P_{k}} {\bf  h^\top} 
\end{aligned}
\end{equation}

% For examples, in general, the sample covariances for a limited number of sample of the innovation sequence is  
%  \begin{equation}
%   \label{eq:t}
%   \begin{aligned}
% E[{\bf \upsilon_{k}} {\bf \upsilon_{k}}^\top]&= \frac{1}{{N_w}-1}\sum\limits_{j=1}^{{N_w}} ({\bf \upsilon_{k-j}}-{\bf\bar \upsilon_{k}}) ({\bf \upsilon_{k-j}}-{\bf\bar \upsilon_{k}}^\top)
%   \end{aligned}
% \end{equation}
According to equation (\ref{eq:24}) and applying the cross-correlation for equation (\ref{eq:23}) and (\ref{eq:24}) at  each epoch. By taking a limited number of sample of the innovation sequence in terms of mean and the covariances are  
 \begin{equation}
  \label{eq:t}
  \begin{aligned}
{\bf\bar \upsilon_{k}}&=\frac{1}{{N_w}}\sum\limits_{j=1}^{{N_w}} {\bf \upsilon_{k-j}},\\
E[{\bf \upsilon_{k}} {\bf \upsilon_{k}}^\top]&= \frac{1}{{N_w}-1}\sum\limits_{j=1}^{{N_w}} ({\bf \upsilon_{k-j}}-{\bf\bar \upsilon_{k}}) ({\bf \upsilon_{k-j}}-{\bf\bar \upsilon_{k}}^\top)
  \end{aligned}
\end{equation}
The combination of innovation and residual sequence is used to improve the Q estimation, in term of  robustness. By considering the sample covariance, thus, the predicted state error covariance is equal to 
 \begin{equation}
 \begin{split}
{\bf \hat P_{k-1}^{-}} &=\frac{1}{a_{1}(k)}\{\frac{1}{2L}\sum\limits_{i=1}^{2L}{({\bf X_{k-1}})_i} {({\bf X_{k-1}})_i^\top} -{\bf \hat x_{k-1}}{\bf \hat x_{k-1}}^{{-}^\top}\} \\& \quad +  {\bf h} {{\bf Q_{k-1}^*}} {\bf  h^\top}
  \label{stateSpaceForm1}
  \end{split}
  \end{equation}
%%%%%%%%%%%%%  for R adaption 
Similarly, we can applying the sample sequence of residual vector as 
   \begin{equation}
{\bf \hat R}_{k-j}^{*}=\frac {1}{N_w}\sum_{j=j0}^{k} {\bf \eta_{j}}{\bf \eta_{j}^\top}-\sum\limits_{i=1}^{2L} {({\bf Z_{k-1}})_i} {({\bf Z_{k-1}})_i^\top}-{\bf \hat z_{k-1}}^{-} {\bf \hat z_{k-1}}^{{-}^\top}
  \label{eq:17}
  \end{equation}
 and thus, the auto covariance of residual sequence ${\bf P_{zz,k}}$ is  evaluated as
\begin{equation} 
{\bf P_{zz,k}}=\sum\limits_{i=0}^{2L}{\bf W_{i}^{c}}[{({\bf Z_k})_i}-{\bf \hat z_{k}}^{-}] [ {({\bf Z_k})_i}-{\bf \hat z_{k}}^{-}]^\top+a_{2}(k){\bf R}_{k}^{*}
 \label{stateSpaceForm1}
 \end{equation}  
  
  The equations (\ref{eq:14}) and (\ref{eq:17}) are the  completes the proof.
  
\end{proof}
% The proposed strategy with a single transitive factor  taking into account the state error covariance adaptively update and also estimate the process noise covariance matrix, Q. 

The pseudo-code of the proposed strategy is given in Algorithm 1.

\begin{algorithm}
  \caption{AFCKF for P and R adaption}\label{euclid}
  \begin{algorithmic}[1]
  %-------------- Input & Output -----------------
  \State \textbf{Input:} Initialize the {$\hat x_{0},\hat P_{0}, \hat Q_{0}, \hat R_{0} $}
  \State \textbf{Compute:} {${{\bf \xi}_i} \gets \sqrt{\bf L}[1]_{i}, {\bf W_{i}}^{m} \gets \frac{1}{2L}  $}
  %--------------- for loop -----------------------
  \For{$k = 1$ to $N$ }
    %   \State {$ S_{k} \gets \ chol ({ \hat P_{k}}) $}
    %   \For {$i = 1$ to $2L+1$ }
 \State{  \textbf{Time update:} \\$ {\bf \hat x_{k-1}}^{-} \gets \frac{1}{2L}\sum\limits_{i=1}^{2L}{({\bf X_{k-1}})_i} $ \\ \textbf{Compute single transitive factor:}\\  $ {\bf a_{1}(k)} = \begin{cases} 1 &\mbox{if } {tr({\bf  P_{zz, k}}}) > {tr({\bf \hat C_{\bf \bar \upsilon k}})} \\ \frac{tr({\bf \hat C_{\bf \bar \upsilon k}})}{tr({ \bf C_{\bf \bar \upsilon k}})} & otherwise \end{cases}$\\
 $  {\bf \hat P_{k-1}^{-}} \gets {\bf a_{1}}(k)( \frac{1}{2L}\sum\limits_{i=1}^{2L}{({\bf X_{k-1}})_i} {({\bf X_{k-1}})_i^\top}-{\bf \hat x_{k-1}}{\bf \hat x_{k-1}}^{{-}^\top})+{\bf Q_{k-1}} $ }
 %     \EndFor
%  \State $ S_{k} \gets \ chol ({\bf \hat P_{k-1}^{-}}) $

% \For {$i \ = \ 1$ to $2L+1$ }
\State{$ {\bf \hat z_{k-1}}^{-} \gets \ \frac{1}{2L}\sum\limits_{i=1}^{2L}{({\bf Z_{k-1}})_i} $ \\ \textbf{Measurement update:}\\ $ {\bf K_{k}}\gets {\bf P_{xz,k-1}}{\bf P_{zz,k-1}^{-1}} $ \\ $ {\bf \hat x_{k}}\gets {\bf \hat x_{k-1}^{-}}+{\bf K_{k}}({\bf z_{k}}-{\bf \hat z_{k-1}^{-}}) $ \\ $ {\bf \hat P_{k}} \gets {\bf \hat P_{k-1}}^{-}-{\bf K_{k}}{\bf P_{zz,k-1}}{\bf K_{k}}^\top $ \\\textbf{ Compute double transitive factor:} \\  $ {\bf a_{1}(k)} = \begin{cases} 1 &\mbox{if } {tr({\bf  P_{zz, k}}}) > {tr({\bf \hat C_{\bf \bar \upsilon k}})} \\ \frac{tr({\bf \hat C_{\bf \bar \upsilon k}})}{tr({ \bf C_{\bf \bar \upsilon k}})} & otherwise \end{cases}$\\  $ {\bf P_{xz,k-1}} \gets \frac{1}{2L}\sum\limits_{i=1}^{2L} {({\bf X_k})_i}  {({\bf Z_k})_i^\top}-{\bf \hat x_{k-1}}^{-} {\bf \hat z_{k-1}}^{{-}^\top} $ \\$ {\bf P_{zz,k-1}}\gets \frac{1}{2L}\sum\limits_{i=1}^{2L} {({\bf Z_{k-1}})_i} {({\bf Z_{k-1}})_i^\top}-{\bf \hat z_{k-1}}^{-} {\bf \hat z_{k-1}}^{{-}^\top}+ {\bf a_{1}(k)} {\bf R_{k-1}^{*}} $ \\
 }
 \EndFor
%\EndFor

  %----------- Remaining text ----------------
 
\State \textbf{Output:} {$\hat x_{k},\hat P_{k}, \hat Q_{k}, \hat P_{k}^{-}, a_{1}(k), a_{2}(k) $ }, 
\end{algorithmic}
\end{algorithm}

The pseudo code for adaptive fading algorithm for ${\bf Q}_{k}^{*}$ and ${\bf R}_{k}^{*}$ is shown in Algorithm 1 and  2, respectively. Note that, we have consider only R-estimation for simulation analysis. 

% In this section for comparing the performance assessment of the proposed algorithm with the existed algorithms with numerical examples in next section.
 
%%
\section{Simulation Results}
Simulation study of a bench-marking target tracking example \cite{huang2020slide} is presented in this section for comparing the performance assessment of the proposed algorithm with the existed algorithms; the CKF, ACKF, AFCKF- P adaption approaches. The nonlinear system and measurement models for target tracking example can be expressed as follows \cite{bar2004estimation,narasimhappa2021}. The state vector, $x_{k}=[ x_{1,k} \quad x_{2,k} \quad x_{3,k} \quad x_{4,k}]^\top$ including the vehicle position and velocity in x and y-plane. $T_s=0.1s$ is the step size. The process, $Q_{k}$ and measurement $R_{k}$  noise covariance matrices are initialized as 

%  \begin{equation}
%  \begin{split}
% {\bf x_{k}} = \begin{bmatrix} 1 & 0 & {T_{s}} & 1 \\ 0 & 1 &  {T_{s}}  & 1 \\ 0 & 0 & 1  &  {T_{s}} \\ 1 & 0 & 1 & {T_{s}} \end{bmatrix} {\bf x_{k-1}}+ \left[ \begin{array}{c} 0 \\ {-k_{s}^2} \\ ({-k_{s}^2} ){T_{s}} \\ ({-k_{s}^2-g}) {T_{s}} \end{array} \right] {\bf w_{k}}
%  \end{split}
%  \end{equation} 
 
%  \begin{equation}
%  \begin{split}
% {\bf z_{1}}(k) = \begin{bmatrix}  \sqrt{(\bf x_{1}({k})-s_{x})^2+(\bf x_{3}(k)-s_{y})^2} \\ {\tan^{-1}( \frac{{\bf x_{3}}(k)-s_{y} }{ {\bf x_{1}}(k)-s_{x}}) } \end{bmatrix} {\bf x_{k}} + {\bf v_{k}}
%  \end{split}
%  \end{equation}

% In this analysis,  initial state, $x_{0}$  = $[0 m,\quad  50 m/s \quad 0 m  \quad 50 m]^\top $  and its state error covariance,  ${\hat P_{0}}$ = diag($[100 m, \quad  100 m/s, \quad 100 m, \quad 100 m/s]$). 

% The nonlinear measurement model is represented as

 \begin{equation}
 \begin{split}
{Q_{0}} =   \begin{bmatrix}
    0 & 0 & 0 & 0 \\
    0 & 2\times10^{-1} & 0 & 0 \\
    0 & 0 & 0 & 0 \\
    0 & 0 & 0 & 2\times10^{-1}
  \end{bmatrix}, {R_{0}} =  \begin{bmatrix}
    100 & 0  \\
    0 & 3\times10^{-4} \\
  \end{bmatrix}
 \end{split}
 \end{equation}

In this section, we consider two different cases for the measurement noises variations as the following: Case A: under Gaussian distribution, Case B: under unknown time-varying measurement noise covariance. The proposed algorithm is compared with other nonlinear cubature filters are implemented including the CKF, ACKF, AFCKF, AFCKF P-adaption. Fig. 1 shows the position estimation of target tracking is obtained. During 4-6 sec, it can be seen that all algorithms can track the actual state, then after the AFCKF-R adaption algorithm has a small deviation from the actual position due to Q value. Subsequently, vehicle seed is better than the other nonlinear approaches. However, the proposed algorithms have better tracking ability in the position estimation when system and measurement model noise change.

% In contrast, in conventional CKF and ACKF,  the state or measurement variations are modeled by a single Gaussian component, whose noise covariance is constant, and wherein ACKF, noise models are varying at each epoch. On the other hand, the performances of the AFCKF-P adaption and AFCKF-R adaption algorithms are better than the CKF, ACKF, and AFCKF methods.  

The RMSE of position and velocity error are shown in Fig. 2 and 3, respectively. It can be seen that the proposed AFCKF R-adaption algorithms yield better estimation accuracy than the non-adaptive CKF and adaptive approaches.  Overall, the proposed AFCKF methods are particularly useful for target tracking state estimation under unknown ambient noises. The average RMSE values of CKF, ACKF and AFCKF algorithms are $1.72 m$, $0.70 m$  and $0.55 m$, respectively.  However, AFCKF-R adaption algorithm outperforms the CKF, ACKF, AFCKF, and AFCKF-P adaption as well. 

% The Root mean square error (RMSE) is a performance indicator and given by
% \begin{equation}
%       \label{eqn4.3}
%       {\bf RMSE_{l}}= \sqrt{\frac{\sum\limits_{i=1}^{N_{sim}} (\bf x_{i}-\bf \hat x_{i})^2}{N_{sim}} }  
%       \end{equation}
% where, $l$ is represents either position or velocity. $N_{sim}$ is the number of  runs. 

% Furthermore, the AFCKF-R algorithm outperforms the CKF and ACKF methods in both estimations, especially for the velocity error as shown on the top-right plot in Fig. 3. 

\begin{table}[htp]
\caption{Average RMSE of considered algorithms }
\label{tab:1}  
\begin{tabular}{lllll}
\cline{2-3}
\hline\noalign{\smallskip}
 &  Case A &  &  Case B  \\
  & RMSE[m] & RMSE[m/s] & RMSE[m] & RMSE[m/s]   \\
%  &  RMSE &  RMSE\\
%   & ${ \degree\per\second}$  & ${\meter\per\second^{2} }$ \\
\noalign{\smallskip}\hline\noalign{\smallskip}
 CKF\cite{arasaratnam2009cubature} &  1.72 &  2.30 &   1.75 &  2.50\\
ACKF\cite{weixi2011multiple} &  0.71 &  1.93&  0.74 &  1.93\\
AFCKF\cite{guo2020robust} &  0.55& 1.02&  0.54 &  0.53 \\
AFCKF-P &  0.52& 0.23&  0.35 &  0.32\\
AFCKF-R &  0.12&  0.21& 0.24 &  0.11\\
\noalign{\smallskip}\hline
\end{tabular}
\end{table}
% Fig. 2 shows the performance of the proposed method in terms of RMSE against time for target tracking. It can be observed that the CKF is having a larger estimation error owing to fixed values of $Q$ and $R$. 

% The average RMSE values of position and velocity errors are reported in Table I. From the table, it can be observed that the average RMSE values of  CKF higher values in both the cases A and B. Even in the CKF, $Q$ and $R$ are constant values. Consequently, ACKF and AFCKF have variation in results because of R is estimated in this case. The proposed algorithms can track close to  the actual state $x_{1,k}$ accurately. Moreover, observed the  RMSE value are very small. 

\begin{figure}[!ht]
\centering
\includegraphics[scale=0.55]{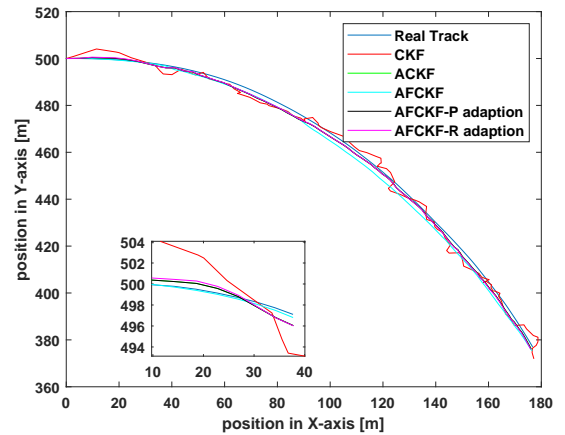}
 \caption{Target tracking result for position and its estimation.}
\label{fig:2}
\end{figure}

\begin{figure}[!ht]
\centering
\includegraphics[scale=0.55]{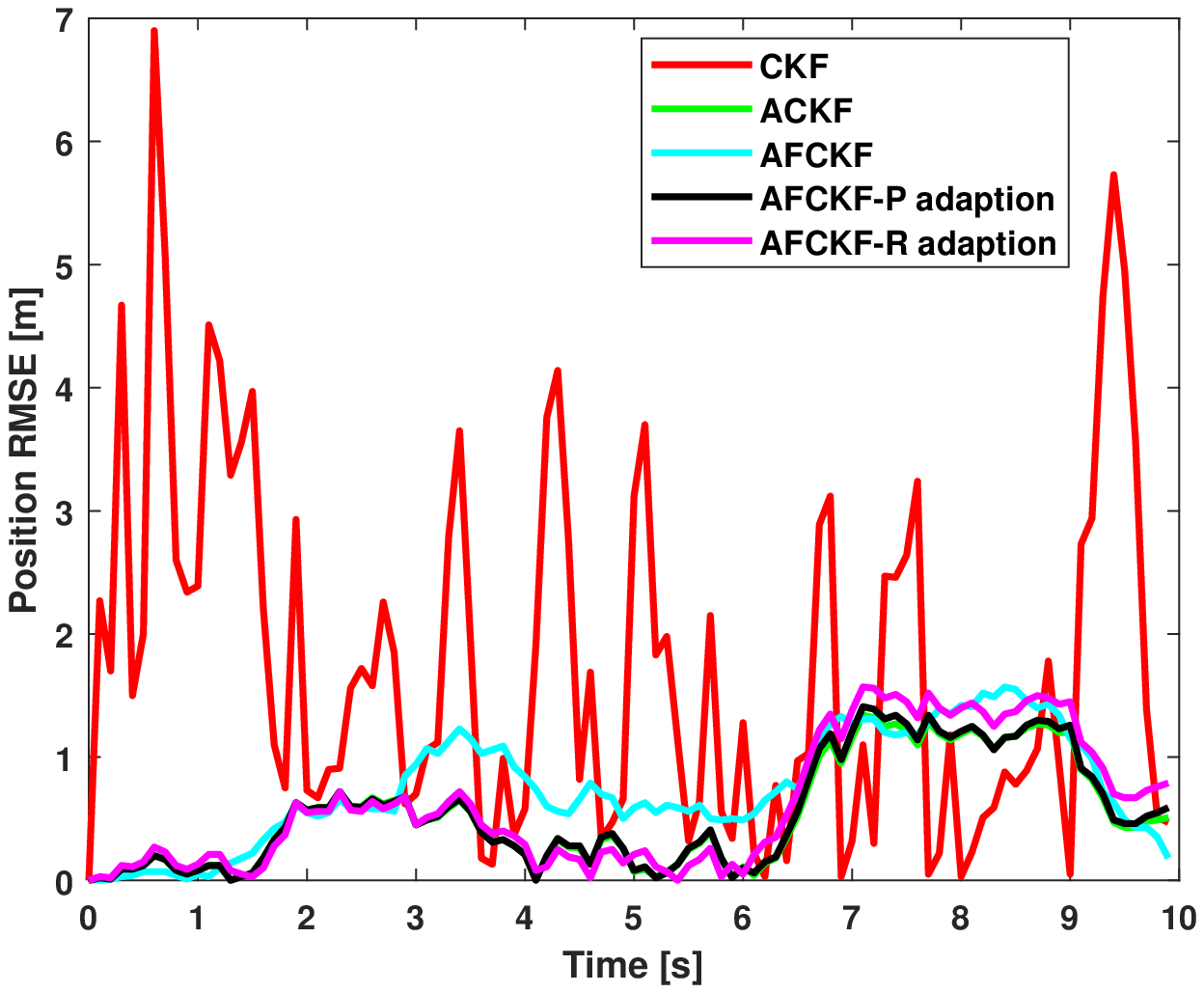}
 \caption{Position RMSE of considered algorithms.}
\label{fig:2}
\end{figure}

\begin{figure}[!ht]
\centering
\includegraphics[scale=0.55]{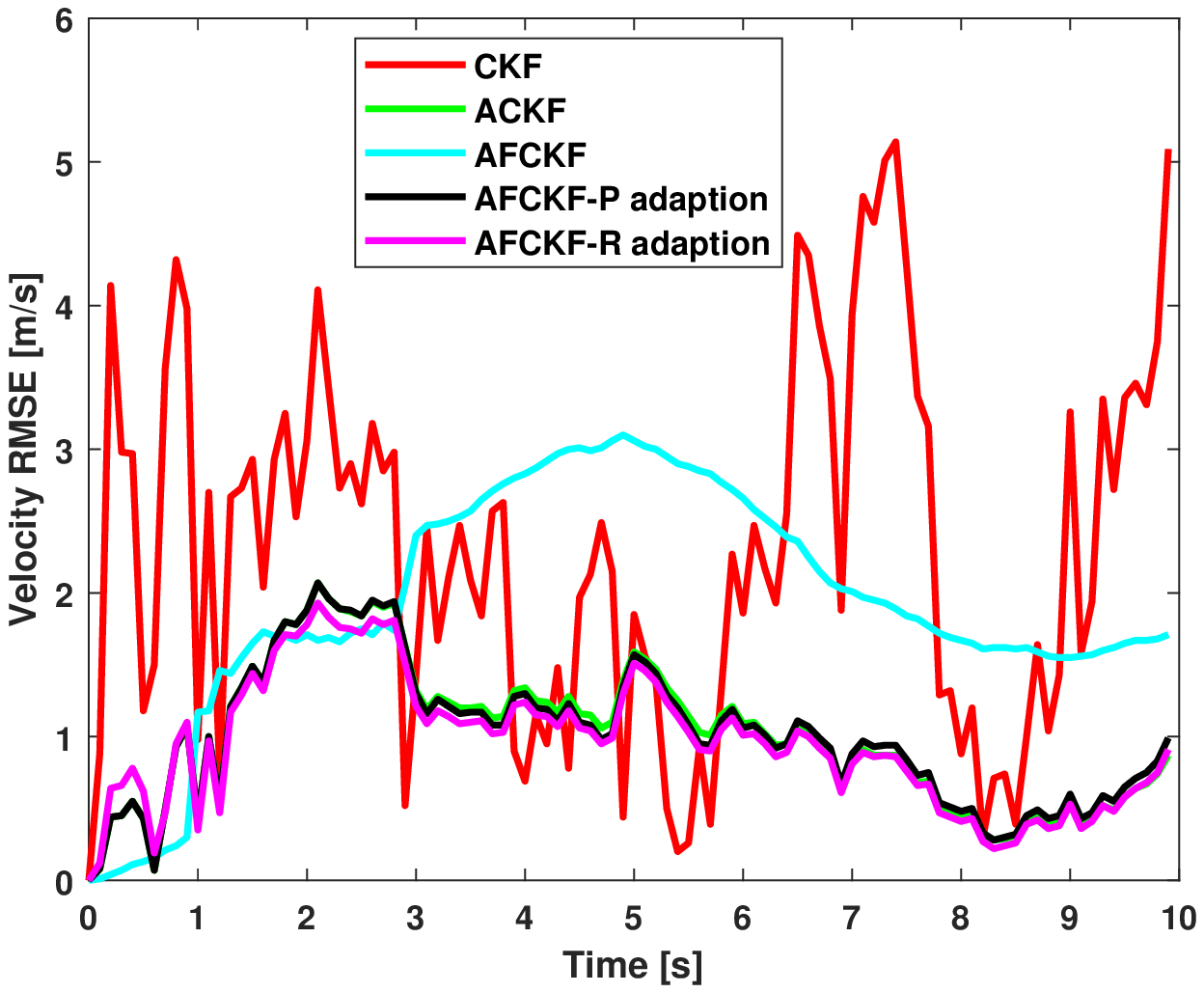}
 \caption{Position RMSE of considered algorithms.}
\label{fig:2}
\end{figure}

The AFCKF-R adaption method can track the true trajectory. Based on the results, it is obvious that two proposed algorithms, AFCKF-P and R adaption methods provide the best results than non-linear filters. The AFCKF-R adaption can obtain better accuracy than the AFCKF-P adaption algorithm under the considerable case A and also more accurate under the time-varying measurement noise, other considerable case B.

% =================================================================================

\section{Conclusion}
The summary of the paper is as follows;
the AFCKF algorithm is developed based on the double transitive factors. In which, the noise covariance matrices are estimated difference between the innovation and  residual sequence. The developed algorithm with ${\bf R_{k}^{*}}$ estimation is only applied in the  application of target tracking . The proposed algorithm could solve the problem of better positioning accuracy, quick converge in the position and velocity, and also to make it better tracking. The RMSE values of considerable algorithms are observed that approximately 20 $\%$ in  position and  30 $\%$ in velocity are improved. Compared with the traditional CKF, ACKF and AFCKF algorithms, the proposed approach has a better adaptability  with the time-varying noise covariance. 
% At the same time, it can also adjust the state error covariance matrix during estimation process with unknown time-varying noise statistics that is fast and accurate.

% , which means it can provide real-time adjustment of the process and  measurement noise matrices.

% \begin{itemize}
% \item The AFCKF algorithm is developed based on the double transitive factors. In which, the combination of  innovation and  residual sequence is used to estimate the process and measurement noise covariance matrices, and also to make it better tracking.
% \item The developed algorithm is applied in the  application of target tracking. The proposed algorithm could solve the problem of better positioning accuracy and observed the quick converge in the position and velocity. At the same time, it can also adjust the state error covariance matrix during estimation process with unknown time-varying noise statistics that is fast and accurate.
% \item The RMSE values of considerable algorithms are observed in this tracking example, we observed that approximately 40 $\%$ in  position and  30 $\%$ in velocity are improved.
% \item  Compared with the traditional CKF, ACKF and AFCKF algorithms, the proposed approach has a better adaptability  with the time-varying noise covariance, which means it can provide real-time adjustment of the process and  measurement noise matrices.
% % \item The simulation result indicated that the proposed algorithm  has  better accuracy than that of existing algorithms, which can effectively improve the estimation precision and inhibit the filtering divergence. 

% \end{itemize}

\bibliographystyle{IEEEtran}
\bibliography{Reference}

\begin{thebibliography}{10}
\providecommand{\url}[1]{#1}
\csname url@samestyle\endcsname
\providecommand{\newblock}{\relax}
\providecommand{\bibinfo}[2]{#2}
\providecommand{\BIBentrySTDinterwordspacing}{\spaceskip=0pt\relax}
\providecommand{\BIBentryALTinterwordstretchfactor}{4}
\providecommand{\BIBentryALTinterwordspacing}{\spaceskip=\fontdimen2\font plus
\BIBentryALTinterwordstretchfactor\fontdimen3\font minus
  \fontdimen4\font\relax}
\providecommand{\BIBforeignlanguage}[2]{{%
\expandafter\ifx\csname l@#1\endcsname\relax
\typeout{** WARNING: IEEEtran.bst: No hyphenation pattern has been}%
\typeout{** loaded for the language `#1'. Using the pattern for}%
\typeout{** the default language instead.}%
\else
\language=\csname l@#1\endcsname
\fi
#2}}
\providecommand{\BIBdecl}{\relax}
\BIBdecl

\bibitem{kim2009stability}
K.-H. Kim, G.-I. Jee, C.-G. Park, and J.-G. Lee, ``The stability analysis of
  the adaptive fading extended {Kalman} filter using the innovation
  covariance,'' \emph{International Journal of Control, Automation and
  Systems}, vol.~7, no.~1, pp. 49--56, 2009.

\bibitem{soken2009adaptive}
H.~E. Soken and C.~Hajiyev, ``Adaptive unscented {Kalman} filter with multiple
  fading factors for pico satellite attitude estimation,'' in \emph{2009 4th
  International Conference on Recent Advances in Space Technologies}.\hskip 1em
  plus 0.5em minus 0.4em\relax IEEE, 2009, pp. 541--546.

\bibitem{arasaratnam2009cubature}
I.~Arasaratnam and S.~Haykin, ``Cubature {Kalman} filters,'' \emph{IEEE
  Transactions on automatic control}, vol.~54, no.~6, pp. 1254--1269, 2009.

\bibitem{guo2020robust}
S.~Guo, L.~Chang, Y.~Li, and Y.~Sun, ``Robust fading cubature {Kalman} filter
  and its application in initial alignment of {SINS},'' \emph{Optik}, vol. 202,
  p. 163593, 2020.

\bibitem{wan2000unscented}
E.~A. Wan and R.~Van Der~Merwe, ``The unscented {Kalman} filter for nonlinear
  estimation,'' in \emph{Proceedings of the IEEE 2000 Adaptive Systems for
  Signal Processing, Communications, and Control Symposium (Cat. No.
  00EX373)}.\hskip 1em plus 0.5em minus 0.4em\relax Ieee, 2000, pp. 153--158.

\bibitem{bar2004estimation}
Y.~Bar-Shalom, X.~R. Li, and T.~Kirubarajan, \emph{Estimation with applications
  to tracking and navigation: theory algorithms and software}.\hskip 1em plus
  0.5em minus 0.4em\relax John Wiley \& Sons, 2004.

\bibitem{yue2019novel}
Z.~Yue, B.~Lian, K.~Tong, and S.~Chen, ``Novel strong tracking square-root
  cubature {Kalman} filter for {GNSS/INS} integrated navigation system,''
  \emph{IET Radar, Sonar \& Navigation}, vol.~13, no.~6, pp. 976--982, 2019.

\bibitem{singh2020exponentially}
A.~K. Singh, ``Exponentially fitted cubature {Kalman} filter with application
  to oscillatory dynamical systems,'' \emph{IEEE Transactions on Circuits and
  Systems I: Regular Papers}, vol.~67, no.~8, pp. 2739--2752, 2020.

\bibitem{li2020application}
S.~Li, Z.~Li, J.~Li, T.~Fernando, H.~H.-C. Iu, Q.~Wang, and X.~Liu,
  ``Application of event-triggered cubature kalman filter for remote nonlinear
  state estimation in wireless sensor network,'' \emph{IEEE Transactions on
  Industrial Electronics}, vol.~68, no.~6, pp. 5133--5145, 2020.

\bibitem{kardan2018improved}
M.~A. Kardan, M.~H. Asemani, A.~Khayatian, N.~Vafamand, M.~H. Khooban,
  T.~Dragi{\v{c}}evi{\'c}, and F.~Blaabjerg, ``Improved stabilization of
  nonlinear {DC} microgrids: Cubature {Kalman} filter approach,'' \emph{IEEE
  Transactions on Industry Applications}, vol.~54, no.~5, pp. 5104--5112, 2018.

\bibitem{pramanik2019accurate}
M.~Pramanik, A.~Routray, and P.~Mitra, ``Accurate real-time estimation of power
  system transients using constrained symmetric strong tracking square-root
  cubature {Kalman} filter,'' \emph{IEEE Access}, vol.~7, pp.
  165\,692--165\,709, 2019.

\bibitem{wang2020generalized}
J.~Wang, Z.~Ma, and X.~Chen, ``Generalized dynamic fuzzy {NN} model based on
  multiple fading factors {SCKF} and its application in integrated
  navigation,'' \emph{IEEE Sensors Journal}, vol.~21, no.~3, pp. 3680--3693,
  2020.

\bibitem{mohamed1999adaptive}
A.~Mohamed and K.~Schwarz, ``Adaptive {Kalman} filtering for {INS/GPS},''
  \emph{Journal of geodesy}, vol.~73, no.~4, pp. 193--203, 1999.

\bibitem{liu2020adaptive}
Z.~Liu and S.-C. Chan, ``Adaptive fading bayesian unscented {Kalman} filter and
  smoother for state estimation of unmanned aircraft systems,'' \emph{IEEE
  Access}, vol.~8, pp. 119\,470--119\,486, 2020.

\bibitem{zhao2015design}
L.~Zhao, J.~Wang, T.~Yu, H.~Jian, and T.~Liu, ``Design of adaptive robust
  square-root cubature {Kalman} filter with noise statistic estimator,''
  \emph{Applied Mathematics and Computation}, vol. 256, pp. 352--367, 2015.

\bibitem{zhou2019new}
W.~Zhou and J.~Hou, ``A new adaptive robust unscented {Kalman} filter for
  improving the accuracy of target tracking,'' \emph{IEEE Access}, vol.~7, pp.
  77\,476--77\,489, 2019.

\bibitem{kim2009adaptive}
K.~H. Kim, J.~G. Lee, and C.~G. Park, ``Adaptive two-stage extended {Kalman}
  filter for a fault-tolerant {INS-GPS} loosely coupled system,'' \emph{IEEE
  Transactions on Aerospace and Electronic Systems}, vol.~45, no.~1, pp.
  125--137, 2009.

\bibitem{xia1994adaptive}
Q.~Xia, M.~Rao, Y.~Ying, and X.~Shen, ``Adaptive fading {Kalman} filter with an
  application,'' \emph{Automatica}, vol.~30, no.~8, pp. 1333--1338, 1994.

\bibitem{weixi2011multiple}
G.~Weixi, M.~Lingjuan, and N.~Maolin, ``Multiple fading factors {Kalman} filter
  for {SINS} static alignment application,'' \emph{Chinese Journal of
  Aeronautics}, vol.~24, no.~4, pp. 476--483, 2011.

\bibitem{li2017stochastic}
L.~Li, D.~Yu, Y.~Xia, and H.~Yang, ``Stochastic stability of a modified
  unscented {Kalman} filter with stochastic nonlinearities and multiple fading
  measurements,'' \emph{Journal of the Franklin Institute}, vol. 354, no.~2,
  pp. 650--667, 2017.

\bibitem{narasimhappa2016arma}
M.~Narasimhappa, J.~Nayak, M.~H. Terra, and S.~L. Sabat, ``{ARMA} model based
  adaptive unscented fading {Kalman} filter for reducing drift of fiber optic
  gyroscope,'' \emph{Sensors and Actuators A: Physical}, vol. 251, pp. 42--51,
  2016.

\bibitem{narasimhappa2021}
M.~Narasimhappa and S.~Srinu, ``Covariance matching based robust adaptive
  cubature {Kalman} filter,'' \emph{arXiv:2106.10775}, pp. 1--5, 2021.

\bibitem{huang2020slide}
Y.~Huang, F.~Zhu, G.~Jia, and Y.~Zhang, ``A slide window variational adaptive
  {Kalman} filter,'' \emph{IEEE Transactions on Circuits and Systems II:
  Express Briefs}, vol.~67, no.~12, pp. 3552--3556, 2020.

\end{thebibliography}

\end{document}